\documentclass{www2009-submission}
\usepackage{url}

\begin{document}
\title{Adding eScience Assets to the Data Web}
%\subtitle{[Extended Abstract]
%\titlenote{A full version of this paper is available as
%\textit{Author's Guide to Preparing ACM SIG Proceedings Using
%\LaTeX$2_\epsilon$\ and BibTeX} at
% You need the command \numberofauthors to handle the "boxing"
% and alignment of the authors under the title, and to add
% a section for authors number 4 through n.
%
% Up to the first three authors are aligned under the title;
% use the \alignauthor commands below to handle those names
% and affiliations. Add names, affiliations, addresses for
% additional authors as the argument to \additionalauthors;
% these will be set for you without further effort on your
% part as the last section in the body of your article BEFORE
% References or any Appendices.

\numberofauthors{3}
%
% Put no more than the first THREE authors in the \author command

% NOTE: All authors should be on the first page. For instructions
% for more than 3 authors, see:
% http://www.acm.org/sigs/pubs/proceed/sigfaq.htm#a18

\author{
%% e-mail address with \email.
\alignauthor Herbert Van de Sompel\\
       \affaddr{Los Alamos National Laboratory}\\
       \affaddr{Los Alamos, NM USA}\\
       \email{herbertv@lanl.gov}
       \alignauthor Carl Lagoze\\
       \affaddr{Cornell University}\\
       \affaddr{Ithaca, NY USA}\\
       \email{lagoze@cs.cornell.edu}
\alignauthor Michael L. Nelson \\
       \affaddr{Old Dominion University} \\
       \affaddr{Norfolk, VA USA}\\
       \email{mln@cs.odu.edu}
\and
\alignauthor Simeon Warner\\
       \affaddr{Cornell University}\\
       \affaddr{Ithaca, NY USA}\\
       \email{simeon@cs.cornell.edu}
\alignauthor Robert Sanderson\\
       \affaddr{University of Liverpool}\\
       \affaddr{Liverpool, UK}\\
       \email{azaroth@liv.ac.uk}
\alignauthor Pete Johnston\\
       \affaddr{Eduserv Foundation}\\
       \affaddr{Bath UK}\\
       \email{pete.johnston@eduserv.org.uk}
}

\maketitle
\begin{abstract}
Aggregations of Web resources are increasingly important in scholarship 
as it adopts new methods that are data-centric, collaborative, and 
networked-based. The same notion of aggregations of resources is 
common to the mashed-up, socially networked information environment of 
Web 2.0.
We present a mechanism to identify and describe aggregations of Web 
resources that has resulted from the Open Archives Initiative -
Object Reuse and Exchange (OAI-ORE) project. The OAI-ORE specifications are based on the principles of the Architecture
of the World Wide Web, the Semantic Web, and the Linked Data effort. Therefore, their incorporation into the 
cyberinfrastructure that supports eScholarship will ensure the integration 
of the products of scholarly research into the Data Web.
%%%% Old version:
%The Open Archives Initiative -
%Object Reuse and Exchange (OAI-ORE) Project defines mechanisms to 
%identify and describe aggregations of Web resources. The aggregations
%are increasingly common in scholarship as it adopts methods that are 
%data-centric, collaborative, and networked-based.  In fact, aggregations 
%of resources are common to the mashed-up, socially networked information 
%environment of Web 2.0.  The OAI-ORE specifications are grounded in Web 
%Architecture principles and the foundations of the Semantic Web and 
%Linked Data.  Therefore, their incorporation into the cyberinfrastructure 
%to support eScholarship will ensure the integration of the products of 
%that endeavor into the general Linked Data Cloud.
\end{abstract}

% A category with only the three required fields



\section{Introduction}

The rapid evolution of computing, networking, and  data capturing
technologies, along with advances in data mining and analysis,
are fundamentally changing the way scholarly research is conducted
\cite{Atkins:2003uq, Borgman:2007fk}.  Although there are differences
amongst disciplines in their receptivity to change \cite{Edwards:2007qy},
an increasing number of scholars in the natural sciences, social sciences,
and humanities have adopted new research methods that are network-based,
highly collaborative, and data-intensive.  Because of the  central role
of vast amounts of data in these new research methods, there has been
increased attention to sustainable infrastructures for registering,
preserving, and sharing datasets \cite{DBLP:journals/corr/cs-DL-0208012}.

In parallel with this change in research methodology there has been
substantial change in the way that research results are communicated. With
the emergence of the Web, scholarly publishers, both commercial
and learned societies, almost universally deliver journal papers,
conference proceedings, and monographs via the Web. While Web delivery
of research results has improved their  accessibility and searchability,
it represents an evolution of traditional publication practices rather
than a fundamental change in the scholarly communication paradigm. Even
in their digital manifestations, scholarly publications are mostly
textually-based and static.  To date, there are few examples of
scholarly communication that move beyond the dissemination of these
traditional artifacts into a more data-centric, semantically-linked, and
social network-embedded scholarly communication model that resembles
the profound changes in social, political, and economic discourse
characteristic of Web 2.0.  This radically different model would expose
process as well as product~\cite{Smith:2008fr}, improving opportunities
to verify the reproducibility of research results, and making the full
spectrum of artifacts generated in the scholarly value chain available
for reuse~\cite{Van-de-Sompel:2004zr}.

The deployment of radically new models depends on the development of
basic technical infrastructure, so-called cyberinfrastructure.  This
cyberinfrastructure must include a number of components.  These include
a means to identify and cite datasets in the scholarly discourse (e.g.,
\cite{sieber:citingdatasets,altman:citing}), a standard
for identifying scholarly authors to unambiguously tie them to their
creations and improve the quality of scientometric information (e.g.,
ResearcherID\footnote{\url{http://www.researcherid.com/}}
and Digital Author
Identifier\footnote{\url{http://www.surffoundation.nl/smartsite.dws?ch=eng&id=13480}}),
and standards to allow machine readability of the products of scholarly
process thereby facilitating computational analysis and extraction
of secondary and tertiary knowledge products. Semantic technologies
are an important component of this cyberinfrastructure,  providing a
foundation for open agreements on data formats, metadata frameworks to
describe data, and ontology-based solutions for formal representation of
scientific knowledge, all of which are important components of promoting
a machine-readable scholarly record.

This paper focuses on one aspect of this cyberinfrastructure that arises
from the changing nature of publications that are characteristic of
collaborative, data-centric scholarship.  These emerging publications are
{\em aggregations} of multiple resources. Such aggregations are already
prevalent in existing scholarly repositories, which commonly offer
access to textual documents in multiple formats, each available from
a different network location. But, the changes in scholarship described
above, and especially the need to include data in the publication process,
increases the complexity of these aggregations and calls for the adoption
of a common approach to handle them. In the remainder of this paper,
we describe our work within Open Archives Initiative - Object Reuse and
Exchange (OAI-ORE), a two-year  project to investigate common methods to
handle aggregations of Web resources that culminated in October 2008 with
the release of the OAI-ORE specifications \cite{Lagoze:2008th}. These
specifications were  motivated by the resource aggregations common
to scholarly communication. We believe that their generic, Web-centric
approach makes them applicable to use cases in the Web at large, providing
the basis for improved search results, improved information navigation,
and richer services within browsers for a large class of Web applications.

The OAI-ORE specifications leverage the principles of the Architecture
of the World Wide Web, the Semantic Web, and the Linked Data effort. As
a result, future developments in cyberinfrastructure and scholarly
communication that are based on OAI-ORE will integrate well with the Web
and with the tools, agents and applications that operate within it. This
will make it possible to embed or mash up the products of scholarship into
cyber-learning efforts, cooperative reference tools such as Wikipedia,
and the larger social discourse that is now characteristic of Web 2.0. The
essence of the OAI-ORE solution to the resource aggregation problem can
be summarized is as follows:

\begin{itemize}
\item The data model is expressed in terms of the primitives of Web Architecture and the Semantic Web: Resources, Representations, URIs and RDF triples.
\item The central entity in the data model, the Aggregation, is a Resource that stands for a set of other Resources. An Aggregation is a Resource with a URI but without a Representation (we refer to this as a non-document Resource from now on).  This approach is aligned with the manner in which real-world entities or concepts are included in the Web via the mechanisms proposed by the Linked Data effort \cite{linkeddata}.
\item	Another Resource, the Resource Map, has a Representation that is a description of the Aggregation. The Resource Map is accessible via the URI of the Aggregation using the mechanisms defined for Cool URIs for the Semantic Web \cite{coolURIssemanticWeb}.
\item The Representation of a Resource Map is a serialization of the triples that describe the Aggregation. The specification describes RDF/XML, RDFa, and Atom serialization syntaxes.
\end{itemize}

\section{Aggregations}
\subsection{Aggregations in Scholarly Communication}

Most institutional repositories~\cite{Johnson:2002mz,Lynch:2003ir}
routinely store and disseminate relatively simple aggregations,
consisting of multiple access formats (e.g., PDF, HTML,
LaTeX) for the same document. In addition, prototypes
exist of applications that allow authoring, storing, and
disseminating more complex scholarly publications in the form of
aggregations~\cite{scope2007,Murray-Rust:2004gf,Williams:2003ve}.
These more complex aggregations may consist of a textual article, one
or more datasets that led to the discoveries reported in the article,
perhaps a visualization of a specific state of the dataset, and the
software used to generate the visualization. All constituents of such an
aggregation are distributed on the Web. One notable aspect of these more
complex visions of an aggregate scholarly publication is the importance
of semantic relationships among constituents of the aggregation. These
relationships include citation, versioning, provenance, commentary,
and the like.

\begin{figure} 
\begin{center}
\includegraphics[scale=0.31]{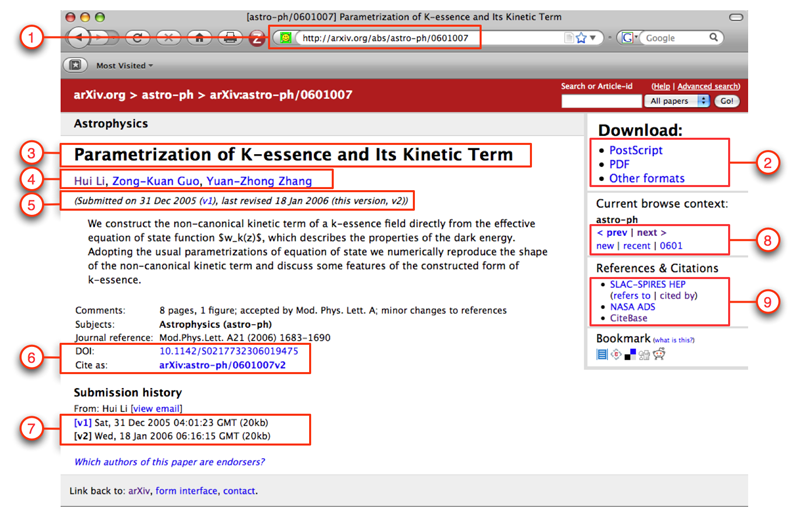}
\caption{The implicitly defined members of a scholarly aggregation.} 
\label{fig:motivating_example}
\end{center}
\end{figure}

Some characteristics of the aggregations that are already
common in scholarship can be illustrated by means of a document
from \url{arXiv.org}, a well-known repository of physics,
mathematics, and computer science research results. The human
start page, or ``splash page'', for this document is shown in Figure
\ref{fig:motivating_example}. Some aspects of the page relevant to the
resource aggregation problem are highlighted in red rectangles, each
with a number. The meanings of the highlighted areas are as follows:

\begin{enumerate} 
\item  The URI \url{http://arxiv.org/abs/astro-ph/0601007} of the human start page for the arXiv document.
\item The formats in which the document is available, i.e. PostScript, PDF, etc. These are effectively the constituents of the aggregation that is the arXiv document. 
\item The title of the arXiv document.
\item The authors of the arXiv document.
\item The creation and last modification date of the arXiv document.
\item Identifiers of resources that are in some manner comparable to this arXiv document. For example, a version of this document was later published as an article in a peer-reviewed journal, and the Digital Object Identifier of that article is shown.
\item The versions of this arXiv document.
\item Links to other arXiv documents in the same collection (i.e., astro-ph).
\item Citations made by this arXiv document, and citations it received from other documents.
\end{enumerate} 

This rather simple example highlights the core issues that OAI-ORE
addresses.  First, although the URI of the human start page is commonly
used as the URI for the entire arXiv document, within the Web Architecture
that URI only identifies the page itself, and not the aggregation that is
the arXiv document.  The ability to cite, annotate, version, and associate
properties with the aggregation itself relies on it having a unique
identity, distinct from the splash page or the resources linked from it.

Second, without the use of (frequently imperfect) heuristics unique
to the specific human start page, it is not readable by machines and
agents. Because the HTML of this human start page usually leaves the
semantics of hyperlinks undefined,  a machine agent cannot unambiguously
distinguish between links to constituents (e.g. the PostScript, PDF,
etc.) of the document and links that point at information that is
clearly outside of the document such as the navigational aids shown as
(8) in Figure \ref{fig:motivating_example}. Similarly, agents can not
interpret relationships of the document to other documents, identifiers
related to this document, versions of this document, etc.

In essence, the problem is that there is no standard way to describe the
constituents or boundary of an aggregation, or to qualify and identify
a resource as being an aggregation.  While a robot could learn the
semantics implied by arXiv's HTML in Figure \ref{fig:motivating_example},
such ``screen scraping'' is brittle and not scalable for applications
accessing aggregations in thousands of different
repositories, each with their own presentation idiom.

\subsection{Integrating Aggregations into the Web}

A number of early efforts in cyberinfrastructure, for example the
initial grid architecture \cite{Tuecke:2003ai} and technologies for
digital libraries, leveraged aspects of the Web infrastructure but often
failed to fully conform with Web Architecture principles. For example,
institutional repositories frequently have identifier schemes and access
protocols distinct from those existing on the Web at large. As a result,
much of their content is accessible on the Web, but it poorly integrates
with mainstream Web applications and may even be overlooked by major
search engines, unless the search engines make special accommodations
for their protocols and access schemes.

Our prior work on the Open Archives Initiative Protocol For Metadata
Harvesting (OAI-PMH)~\cite{379449} demonstrates this
problem. OAI-PMH is an interoperability specification released in 2001
aimed at streamlining the process of incrementally collecting XML metadata
(typically bibliographic metadata) from information systems. It shares
many design characteristics with Atom~\cite{rfc:4287} and is widely
adopted in its targeted community of scholarly repositories. But, OAI-PMH,
in contrast to Atom, has not gained broader adoption, mainly because its
architecture is not well aligned with the Resource/URI/Representation
foundations of the Web Architecture. For example, OAI-PMH clients
must construct a request URI by combining a repository specific base
URI, the identifier of the item of interest, and a format tag in an
OAI-PMH specific manner, often preventing general Web clients that
are unaware of the protocol from accessing the available metadata
\cite{haslhofer2008ose}.

The Web-centric, resource-centric approach of OAI-ORE rectifies this architectural shortcoming and thereby provides the foundation for full accessibility of
the products of eScience in the general Web environment. Furthermore,
it makes the solution available to a broader class of Web
applications in which the practice of aggregating resources is quite
common. For example, we accumulate URLs in bookmarks or favorites lists
in our browser, collect photos into sets in popular sites like Flickr,
browse over multiple page documents that are linked together through
``prev'' and ``next'' tags, and talk about Web sites as if they had some
real existence beyond the set of pages of which they consist. Despite
our frequent use of these aggregations, their existence on the Web is
quite ephemeral because there is no common way to identify, describe,
and hence handle them.  This is what OAI-ORE provides.

\section{The OAI-ORE Solution}

In this section we describe the various elements of the OAI-ORE solution
to the resource aggregation problem outlined above. It encompasses an RDF-based
data model, syntaxes for serializing instances of the data model, and
mechanisms for providing HTTP access to those serializations. Complete
details are available through the OAI-ORE documentation suite
\cite{Lagoze:2008th}.

As noted earlier, this solution is based on the primitives defined
in the Architecture of the World Wide Web \cite{archWWW} that defines
a Resource as an item of interest; a URI as a global identifier for
a Resource; and a Representation as a datastream corresponding to the
state of a Resource at the time its URI is dereferenced via some protocol
(e.g. HTTP).  In addition, the solution is grounded in the principles
introduced by the Semantic Web, in which URIs are also used to identify
non-document Resources, such as real-world entities (e.g. people or cars),
or even abstract entities (e.g. ideas or classes).  These non-document
Resources have no Representation to indicate their meaning. OAI-ORE adopts the following approach, proposed
by the
Linked Data effort \cite{linkeddata}, for obtaining information about those Resources:

\begin{itemize}
\item Use of HTTP URIs to identify those non-document Resources;
\item Publication of another Resource with a Representation that provides information about the non-document Resource at a HTTP URI other than the HTTP URI of the non-document Resource;
%\item Use of Cool URIs for the Semantic Web~\cite{coolURIssemanticWeb} to allow discovery of the HTTP URI of that other Resource from the HTTP URI of the non-document Resource.
\item Leverage of HTTP mechanisms to allow discovery of the HTTP URI of the published resource from the HTTP URI of the non-document resource.
\end{itemize}

\subsection{Data Model}

\begin{figure*}
\begin{center}
\includegraphics[scale=0.55]{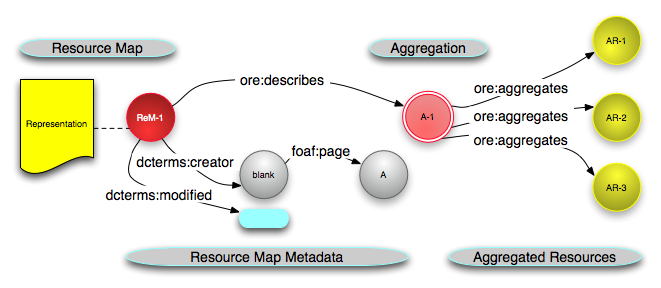}
\caption{A Resource Map describes an Aggregation with three Aggregated Resources.}
\label{fig:ADM_basics}
\end{center}
\end{figure*}

The essence of the RDF-based data model is described here and is illustrated in 
Figure~\ref{fig:ADM_basics}. The full details are available in the OAI-ORE Abstract 
Data Model specification~\cite{ore:datamodel}.

In order to be able to unambiguously refer to an aggregation of
Web resources, a new Resource is introduced that stands for a
set or collection of other Resources. This new Resource, named an
{\em Aggregation}, has a URI just like any other Resource on the Web. 
And, since an Aggregation is a conceptual construct, it is 
a non-document Resource that does not have a Representation.

Following the Linked Data guidelines, another Resource is introduced to
make information about the Aggregation available. This new Resource, named
a {\em Resource Map}, has a URI and a machine-readable Representation
that provides details about the Aggregation. In essence, a Resource Map
expresses which Aggregation it describes (the \texttt{ore:describes}
relationship in Figure \ref{fig:ADM_basics}), and it lists the
{\em Aggregated Resources} that are part of the Aggregation (the
\texttt{ore:aggregates} relationship in Figure \ref{fig:ADM_basics},
a subproperty of \\
\texttt{dcterms:hasPart}). But, a Resource Map can
also express relationships and properties pertaining to all these
Resources, as well as metadata pertaining to the Resource Map itself,
e.g. who published it and when it was most recently modified (the
\texttt{dcterms:creator} and \texttt{dcterms:modified} relationships
in Figure \ref{fig:ADM_basics}). A Resource Map can also express
relationships of the Aggregation, Aggregated Resources, and the Resource
Map itself with any arbitrary other Resource, as long as the resulting
RDF graph is connected.

In addition, for discovery purposes, the data model allows a
Resource Map to express that an Aggregated Resource of a specific
Aggregation is also part of another Aggregation. This is achieved by
means of the \texttt{ore:isAggregatedBy} relationship (the inverse
of \texttt{ore:aggregates}) between the Aggregated Resource and
that other Aggregation. Also stating that an Aggregated Resource is
itself an Aggregation (nesting Aggregations) is supported. To that
purpose, an \texttt{ore:isDescribedBy} relationship (the inverse of\\
\texttt{ore:describes}, and a subproperty of \texttt{rdfs:seeAlso})
is expressed between the Aggregated Resource and a Resource Map
that describes it as being itself an Aggregation. Furthermore,
the use of non-protocol-based identifiers (such as DOIs) that
can be expressed as URIs is quite common for referencing scholarly
assets. In order to support this practice, the \texttt{ore:similarTo}
relationship between an Aggregation and a somehow equivalent resource
identified by a non-protocol-based URI is expressed. The specificity
of \texttt{ore:similarTo} is situated between \texttt{rdfs:seeAlso}
and \\
\texttt{owl:sameAs}.

\subsection{Proxies: Aggregated Resources in Context}

We note that the URI asserted in a Resource Map to denote an Aggregated
Resource of a particular Aggregation is no different than the URI that
denotes that Resource independent of the Aggregation.  However, it is
important in scholarly communication, among others for the purpose of
citing and expressing provenance, that a resource such as a dataset
included in some context, for example a specific article, be distinct
from the same dataset outside the context of that article, or in the
context of another article.

\begin{figure}
\begin{center}
\includegraphics[scale=0.28]{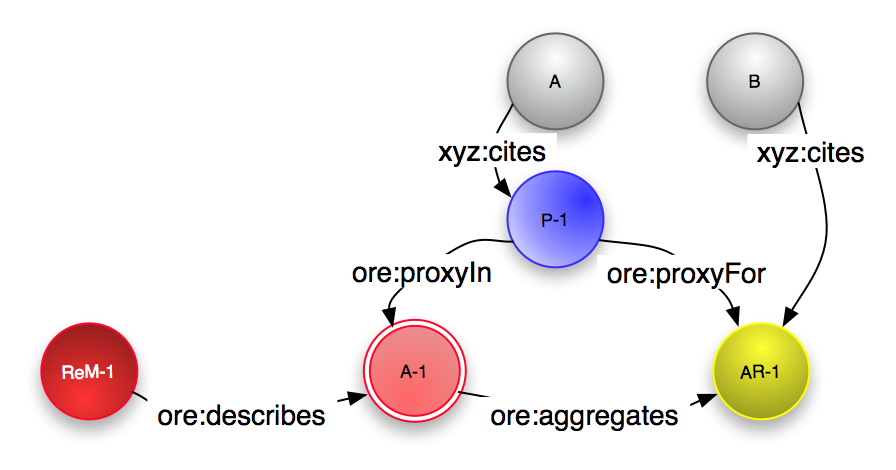}
\caption{Citing a Resource in the context of an Aggregation.}
\label{fig:ContextLink}
\end{center}
\end{figure}

To accomplish this differentiation, OAI-ORE introduces the notion of
a {\em Proxy}. A Proxy is a Resource that stands for an Aggregated Resource
in the context of a specific Aggregation. The URI of a Proxy provides a
mechanism for denoting a Resource in context. Figure \ref{fig:ContextLink}
shows the \texttt{ore:ProxyFor} and \texttt{ore:ProxyIn} relationships
between a Proxy and an Aggregated Resource and an Aggregation,
respectively. It also illustrates how citing the Aggregated Resource is
different from citing its Proxy: the former cites a Resource ``as is'',
the latter cites that Resource as it exists in the context of a specific
Aggregation. In order to work seamlessly in the Web and to provide 
context information to OAI-ORE aware clients, resolution of HTTP URIs 
assigned to Proxies must lead to the Aggregated Resource, and the 
response must include a HTTP Link Header~\cite{rfc:http-link} that 
points to the Aggregation.

\subsection{Resource Map Serializations}

A Resource Map has a Representation that describes an Aggregation in some
serialization syntax. OAI-ORE explicitly specifies three serialization
syntaxes, Atom XML, RDF/XML, and RDFa, while other serialization
syntaxes are possible. Which one to choose will largely depend on the
use case and on the technical environment available to a Resource Map
publisher. For example, in cases where an expressive HTML splash page
exists an RDFa approach might be attractive. Note that multiple Resource
Maps, each using a different serialization syntax can describe the same
Aggregation, and that these may differ in expressiveness\footnote{See 
\url{http://www.openarchives.org/ore/atom} for detailed Atom 
and RDF/XML versions of Resources Maps corresponding to Figure \ref{fig:motivating_example}.}.

Although the data model is based on RDF, we were committed
to also specify a serialization based on Atom, to allow Aggregations
to become the subject of Web 2.0 reuse scenarios and of workflows
based on the Atom Publishing Protocol \cite{rfc:5023}. The Atom
Publishing Protocol adds a uniform read/write approach to Web 2.0,
which could be of significant benefit in scholarly communication
scenarios. 

However, the task of reconciling the data model with the Atom model
proved to be non-trivial due to tensions between the RDF model and the
XML-oriented Atom specification.  The former is graph-based, with precise
semantics that are global rather than local to a specific document.
The latter is hierarchical, (XML) document-centric, and has intentionally
loose element definitions. It took several, dramatically different
iterations of the Atom serialization to arrive at an acceptable solution.

The resulting approach expresses an Aggregation by means of an
Atom \texttt{entry}, and makes use of Atom's extensibility mechanisms in
much the same way as Google Data does. For example, Atom's \texttt{link}
element with an OAI-ORE-specific value for the \texttt{rel} attribute
is used to aggregate resources. And, awaiting a solution from the Atom
community to deal express triples, an \texttt{ore:triples} element
was introduced to act as a wrapper for RDF descriptions. To support
unambiguous interpretation of Atom serializations of Resource Maps, a
GRDDL transform was implemented that extracts all contained triples that
pertain to the OAI-ORE data model, both from the native Atom elements
and from the \texttt{ore:triples} extension element, and expresses them
in RDF/XML\footnote{http://www.openarchives.org/ore/atom-grddl}.

\subsection{Leveraging HTTP}

\begin{figure}
\begin{center}
\includegraphics[scale=0.50]{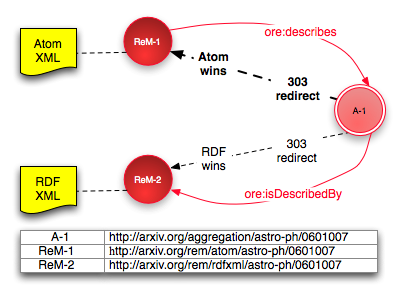}
\caption{Discovering a Resource Map from an Aggregation using Cool URIs for the Semantic Web.}
\label{fig:RedirectCN_1}
\end{center}
\end{figure}

In order to make OAI-ORE work in the HTTP-based Web, both the Aggregation
and the Resource Map are assigned HTTP URIs, and the Cool URIs for
the Semantic Web guidelines \cite{coolURIssemanticWeb} are adopted to
support discovery of the HTTP URI of a Resource Map given the HTTP URI
of an Aggregation. Figure \ref{fig:RedirectCN_1} illustrates a situation
in which the arXiv Aggregation is described by both an Atom XML and an
RDF/XML Resource Map, and in which a client is led to the Atom version
via an HTTP 303 redirect and Content Negotiation.

\subsection{Authoritative Resource Maps}

After one party has published a Resource Map that contains a
description and a URI for a new Aggregation, any other party can
publish competing or even conflicting Resource Maps that describe the
same Aggregation. To address this we distinguish between 
Authoritative and Non-Authoritative Resource Maps in the same way
as the Linked Data guidelines. An
Authoritative Resource Map is one that is accessible by dereferencing the URI of
the Aggregation that it describes, for example using the aforementioned
Cool URI mechanisms. A Non-Authoritative Resource Map is one not reachable
in this manner. The rationale for this approach is that the party that
introduces a new Aggregation simultaneously mints URIs for both the
Aggregation and the Resource Map, and actually controls both.

\section{Early Demonstrators}

Since the OAI-ORE specifications have only been released recently,
an in-depth evaluation of functionality, adoption, and impact is
premature. Still, in this section we give an insight in efforts
by early adopters to leverage the specifications.  Four use cases
are described below. Additional illustrations of its application
are provided by the submissions to the ORE Challenge at RepoCamp
2008\footnote{\url{http://www.openarchives.org/ore/RepoCamp2008/}}.

\subsection{Foresite: Revealing Aggregations}

In order to provide feedback on the evolving OAI-ORE specification, the UK's Joint Information
Systems Committee (JISC)\footnote{\url{http://www.jisc.ac.uk/}}
funded an experiment to investigate applying it to an extensive
scholarly collection: the approximately four million articles that
are part of the JSTOR\footnote{\url{http://www.jstor.org/}}
collection. By developing open source OAI-ORE
libraries\footnote{\url{http://foresite-toolkit.googlecode.com/}} and
applying them to produce interlinked Resource Maps, the Foresite project
effectively demonstrated the feasibility of exposing common scholarly
artifacts to the Data Web in the manner proposed by OAI-ORE. The
project provided valuable feedback that helped refine the OAI-ORE
specifications, and had a significant impact on the aforementioned
discussions regarding the Atom serialization of Resource Maps.

\begin{figure}
\begin{center}
\includegraphics[scale=0.45]{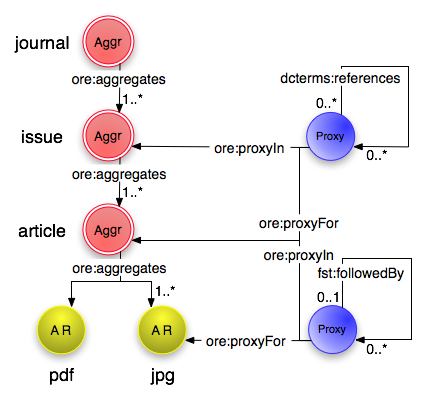}
\caption{The hierarchical structure of the JSTOR collection mapped to the OAI-ORE data model. Note that \texttt{1..1} cardinalities are omitted from the diagram for clarity.}
\label{fig:jstor-uml}
\end{center}
\end{figure}

The overall structure of the Aggregations, and associated Resource Maps,
produced for the JSTOR collection mirrors the journal - issue - article
hierarchy of the JSTOR content. Each journal is modeled as an Aggregation
of journal issues; each issue is an Aggregation of articles; and each
article is an Aggregation of individual page images and a PDF-formatted
version of the entire article (Figure \ref{fig:jstor-uml}). The
Aggregated Resources at each level are also the subject and/or object
of a \texttt{fst:followedBy} relationship introduced to preserve the
page-turning order for pages within an article, articles within an
issue and so forth. Because \texttt{fst:followedBy} is not a global
relationship, but rather only applies within the context of a specific
Aggregation, Proxies for these Aggregated Resources were introduced. The
article Aggregations interlink via \texttt{dcterms:references}
relationships for citations, further confirming the necessity of the
graph-based nature of the OAI-ORE date model, even though the main JSTOR
content hierarchy is tree-shaped. The Resource Maps were published on
a Web server at the University of Liverpool.

The resulting OAI-ORE descriptions are of immediate business importance
to JSTOR. While JSTOR stores the OCR-ed full-text of each article, it
is only able to openly expose this kind of topological metadata, and
would lose its market advantage (and the participation of contributing
publishers) if the full-text were exposed. Having the topology of their
collection available in a standardized format that provides links back
to their protected full-text documents and images, facilitates reuse in
third party applications that can help drive traffic to the JSTOR site
and increase its customer base.

In order to provide a value-added service on the basis of the generated
Resource Maps without requiring JSTOR to integrate prototype code
into their production portal, the Foresite Explorer -- a visualization
application\footnote{ \url{http://foresite.cheshire3.org/explorer/}}, was
developed using GreaseMonkey\footnote{\url{http://www.greasespot.net/}}
and its cross-site capable XmlHttpRequest. This one-click-install plug-in
for Firefox\footnote{\url{http://www.mozilla.com/firefox/}} extracts
the URI of the resource that is currently being viewed in the JSTOR
Web interface and retrieves the associated RDF/XML Resource Map that
describes the Aggregation to which the Web resource corresponds from the
Liverpool Web server. The plug-in then parses and displays the Resource
Map graph via dynamic SVG. Nodes in the display represent Aggregations,
Aggregated Resources, and related Resources. Nodes for Aggregations can
be clicked to expand or contract the visualization; in case of expansion,
new Resource Maps are obtained, parsed, and again visualized.

\begin{figure}
\begin{center}
\includegraphics[scale=0.25]{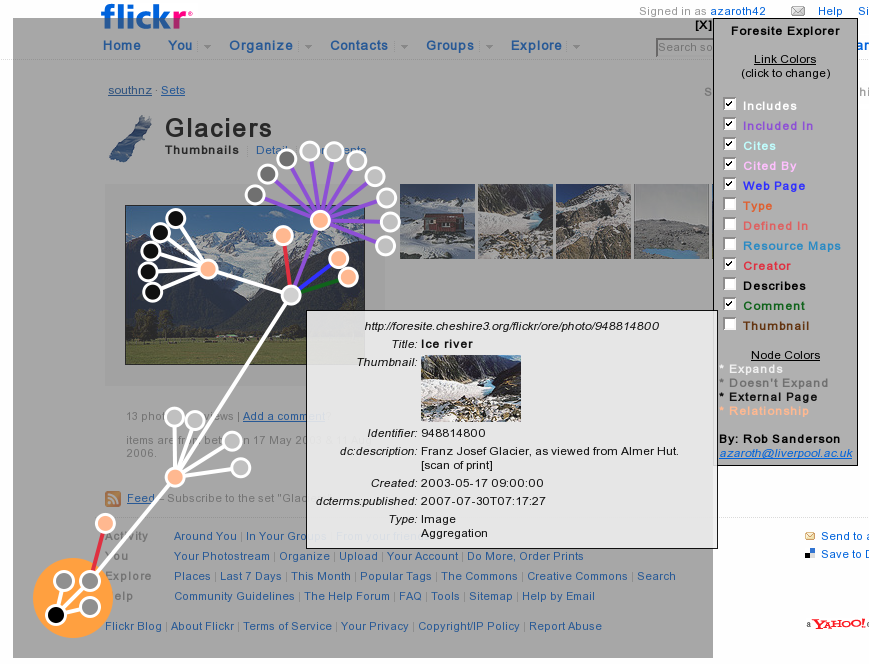}
\caption{The Foresite plug-in models Flickr Sets as OAI-ORE Aggregations, and visualizes them.}
\label{fig:Flickr}
\end{center}
\end{figure}

Further experiments using the same approach were carried out
on mainstream Web portals, leveraging the provided Web service
APIs to obtain metadata, and to express it according to the
ORE data model. Flickr\footnote{\url{http://www.flickr.com/}} and
Amazon\footnote{\url{http://www.amazon.com/}} were selected, and wrapper
services were built to generate Resource Maps on demand through REST
interactions, and to publish them on the Liverpool server. Flickr provides
a rich dataset with photos, photo sets, users, groups, favorites and
even comments and tags that can all be modeled as Aggregations. Figure
\ref{fig:Flickr} shows a visualization of the structure of the Flickr
Set ``Glaciers'' that consists of five photographs. In the Foresite
Explorer, this set is represented with an Aggregation visualized as
the top right node within the OAI-ORE logo (left bottom of Figure
\ref{fig:Flickr}), emitting a red \texttt{dcterms:creator} arc and
a white \texttt{ore:aggregates} arc. The latter leads to the five
photographs.  The third photograph is selected, and another white
\texttt{ore:aggregates} arc reaches out to the available image files
(differing image resolutions) represented as black nodes. The purple nodes
indicate other aggregations in which the selected photo is aggregated.

Amazon offers fewer constructs that readily map to the OAI-ORE data model,
but the user wishlists is a compelling one.  The mapping to the data 
model is as follows: a wishlist becomes an Aggregation, and wished-for
items become Aggregated Resources. Interestingly, each item in an
Amazon wishlist has a unique identifier by which it is purchased. That
identifier is only valid within that specific wishlist to allow tracking
of individual items, once purchased. These wishlist specific constructs
map directly the Proxies of the OAI-ORE model.  The GreaseMonkey script
was updated to discover these identifiers that are necessary to interact
with the Amazon Web services, and Proxy-based relationships were added
to the visualization.

Overall, the Foresite experiment has illustrated the applicability of the
OAI-ORE resource aggregation model as well as the feasibility to leverage
it to create a value-added service. It has demonstrated this for both
common scholarly communication artifacts and specific constructs used
by popular Web portals. The Foresite experiment will be described in
more detail in a dedicated, future publication.

\subsection{Astronomy Publication Workflow}

Datasets are of fundamental importance in observational sciences such
as astronomy. The astronomy community has developed sophisticated
repositories and data standards, exemplified by the Sloan Digital Sky
Survey\footnote{\url{http://www.sdss.org/}} and the National Virtual
Observatory\footnote{\url{http://www.us-vo.org/}}, which provide excellent
facilities for registering and accessing large datasets. However, when
submitting an article, both new datasets that were created to arrive
at findings reported in an article, and data citation information that
reveals the reuse of existing datasets are often lost, ``left behind''
on the personal computer of the author.

A team at Johns Hopkins University is collaborating with the American
Astronomical Society to capture datasets as part of the publication
workflow~\cite{choudhury2007}. In the newly devised publication workflows,
OAI-ORE Aggregations are used to glue an article and its associated
datasets together, and Resource Maps that describe these Aggregations
are the tokens that move around between author, publisher and dataset
repository as the publication process proceeds~\cite{dilauro2008}. At
each stage of the publication workflow, the Resource Map is used to
convey the current state of the Aggregation, and is then updated to
reflect the new state that is then passed on to the next workflow
phase. For example, as a Resource Map is passed from the publisher to
the dataset repository and back again, it is updated to contain the URIs
of datasets that are registered in the repository, and that were used
for the article. This allows the publisher to link to the datasets that
were used for a specific article, and the repository to link to papers
that used a specific dataset.

Generally, the availability of these Aggregations enables new services to be built on both the publishing platform and the data repository. If the practices proposed by this novel publication workflow became commonplace, it would represent a significant improvement in the efficiency of scientific communication.

\subsection{Authoring, Editing and Reusing}

The success of OAI-ORE depends on the ease with which Aggregations
and Resource Maps are authored and disseminated on the Web.  In many
cases, they will be generated automatically based on information that
is available in an information system. For example, the \url{arXiv.org}
database contains all information that is necessary to automatically
generate Aggregations and their associated Resource Maps, as shown in the
Appendices. And, in the  astronomy project described above, the ability
to create Resource Maps is built into familiar authoring environments
in a manner that makes it a side-effect of the authoring process and
thus minimizes the burden on authors.

Like all cyberinfrastructure, the success of such authoring environments
depends on the manner in which assembling all resources that relate to a
particular research task or publication fits into the normal scholarly
workflow. Two authoring environments that demonstrate this are the
Literature Object Reuse and Exchange (LORE) tool created by Gerber et
al.\footnote{\url{http://www.openarchives.org/ore/RepoCamp2008/#LORE}},
and by the SCOPE work of Cheung et al.~\cite{scope2007,hunter2007}.
LORE is a Firefox extension that communicates via Ajax with a Sesame2
data store for maintaining the OAI-ORE graphs that are generated.
LORE allows for the generation of fine-grained metadata and relationships,
for example, allowing indicating that a certain resource is contextual
information about the literature work that is being studied. The SCOPE
work led to the development of the Provenance Explorer, a stand-alone
Java application with functionalities similar to those of LORE, but aimed
at the creation, editing and publication of scientific compound objects.

\subsection{Enhanced Publications}

The Dutch SURFshare
program\footnote{\url{http://www.surffoundation.nl/en/}} and the European \\
DRIVER II project\footnote{\url{http://www.driver-community.eu/}} are
collaborating on cyberinfrastructure to join a multitude of scientific
repositories that hold publications and research data. The goal is to
give researchers better means to share and access scientific materials
through innovative services. One of the envisioned services relates to
\textit{enhanced publications}, composites of textual publications and
supporting resources such as research-data, visualizations, annotations,
related websites, etc. To ensure the integrity and usability of such
enhanced publications it is important that all its components and their
interrelations are being preserved.

\begin{figure}
\begin{center}
\includegraphics[scale=0.12]{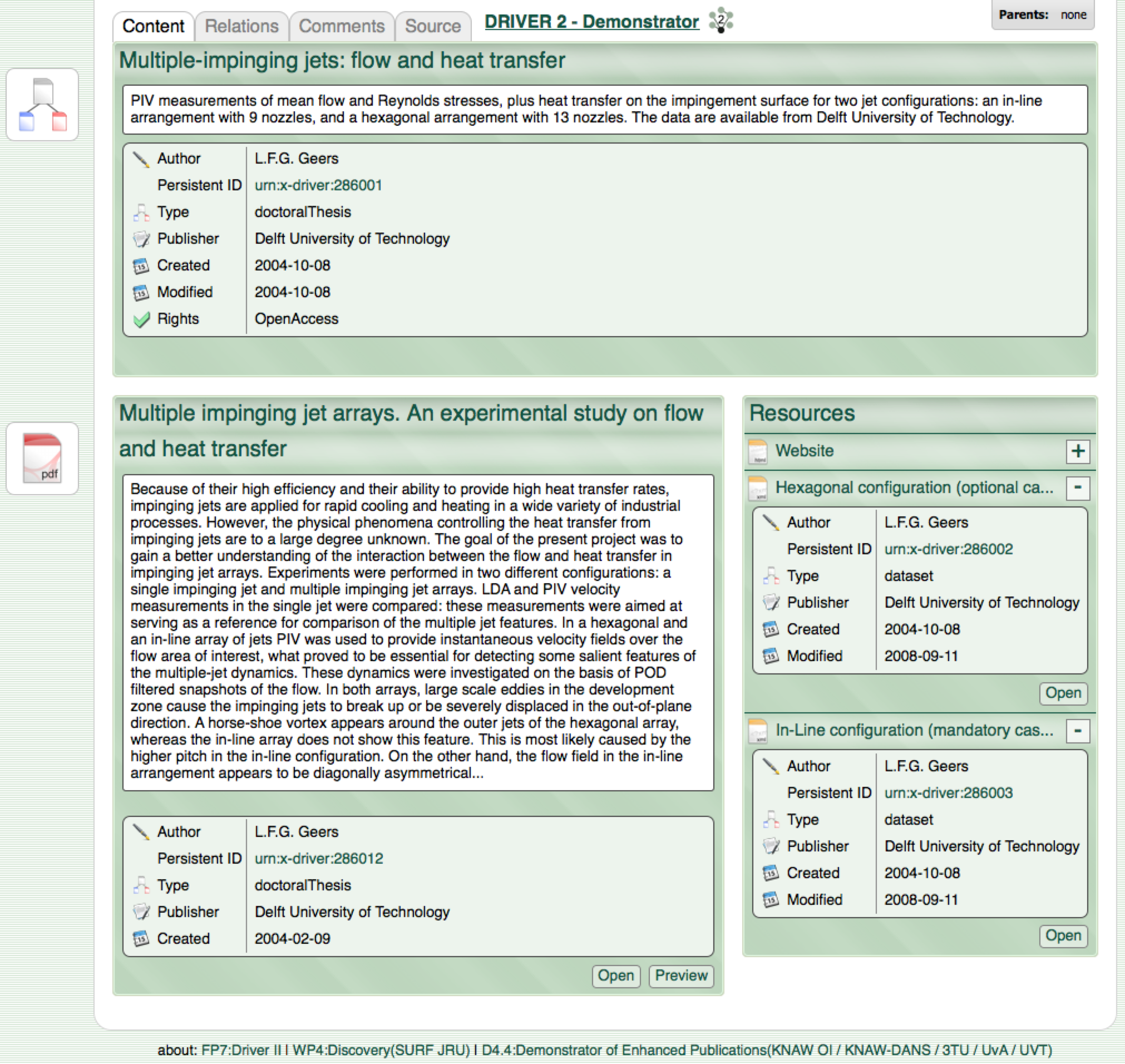}
\caption{The splash page for an enhanced publication of the DRIVER II project, dynamically
rendered from an RDF/XML Resource Map.}
\label{fig:driver2}
\end{center}
\end{figure}

A study into object models suitable for the representation of enhanced
publications recommended the use of OAI-ORE.  As a result, a demonstrator
project \cite{driver2} was launched in which enhanced publications for multiple
scientific disciplines ranging from engineering to journalism were
modeled according to OAI-ORE, and in which approaches to meet a variety
of requirements were explored, including presentation, navigation,
persistent identification, granularity of referencing, handling of
sequentially ordered resources, visualization of interrelationships,
etc. The results are available at the project site\footnote{\url{
http://driver2.dans.knaw.nl/demonstrator/html/}}. The project chose
RDF/XML to express Resource Maps and uses an XSLT-based approach to
dynamically generate an HTML ``splash page'' from them. In each splash
page, a \textit{Content} tab (Figure \ref{fig:driver2}) lists all crucial
metadata about the enhanced publication, prominently shows its textual
component and associated metadata, and neatly lists additional resources
again with metadata. Many of these resources are themselves modeled as
Aggregations, and hence also have their own splash page. To support an
understanding of the relationships among resources of an Aggregation
and of nested Aggregations, a \textit{Relations} tab that loads a
Java applet fueled by Resource Map content is introduced. Overall,
the demonstrator is remarkable because of the elegance and simplicity
of the ORE implementation. It clearly illustrates that ORE can be used
as a basic model for enhanced publications, and points at the need for
community-defined vocabularies to convey expressive relationships among
scientific resources.

\section{Related Work}

Given the widespread use of aggregations in both the physical and the Web
world, it comes as no surprise that other efforts have investigated this
domain. Prior work in the Web realm can be grouped in two main categories
depending on the party that introduces aggregations. In one case, that
is the Web navigator (agent or reader), in the other case it is the
administrator of a Web-based information system. We look at a number of
efforts in both categories, and evaluate their capabilities to identify
aggregations, to enumerate the constituent resources of an aggregation,
to express relationships among resources, and to accommodate resources
that are distributed on the Web.

In the Web navigator case, either an interactive user groups resources
based on some intent, or a robot tries to infer the implicitly defined
members of an aggregation. The robotic approaches range from heuristics
\cite{li2000dld, eiron2003ucd} to machine-learning \cite{dmitriev2005wmp,
dmitriev2008wmp}.  While these approaches are useful, they are imperfect
and dependent on the perception of those encoding the heuristics or
training set and they do not necessarily reflect the intention of the
original authors of the Web resources.  And, while these approaches
may succeed at selecting the distributed resources that are part of an
implicitly defined aggregation, they are not capable of inferring the
relationships between those resources, nor do they propose a way to
unambiguously describe the aggregation.

The approaches that involve an interactive user include
tools such as GroupMe!\footnote{\url{http://groupme.org/}} and
LinkBunch\footnote{\url{http://linkbun.ch/}}. LinkBunch lets users
submit several URIs that are then assigned a new HTTP URI that, when
dereferenced, returns an HTML page that lists and links to the originally
submitted URIs.  The ``bunch'' has a new HTTP URI identity, it enumerates
its members, and it readily handles distributed Web resources. However,
the identity of the bunch is the same as that of the HTML page that
describes it, and expressing relationships between the bunched resources
is not supported.  GroupMe! is similar, with the addition of social
tagging capabilities, but has the same problems as LinkBunch.

Some Web navigator approaches work in an opposite granular direction,
supporting \textit{disaggregation} of a single Web resource (i.e., an
HTML page) into multiple resources. This can be done automatically,
such as for segmented display on limited devices such as PDAs
\cite{chakrabarti2008gta} or for recovering structured records
from Web pages \cite{embley1999rbd}. Decomposition can also be done
manually, such as for reuse and sharing of parts of a Web page (e.g.,
ClipMarks\footnote{\url{http://clipmarks.com/}}).  All these approaches,
manually or automatically, can be thought of as adding (or inferring) HTML
anchors where none exist.  These approaches assign identities to the newly
created resources (fragments of the original resource), but they provide
no approach to describe the original resource as an aggregation of these
new resources, nor do they allow expressing relationships among them.

In approaches that have the administrator of a Web information system
in the diver seat, several technologies exist to deal with resource
aggregations. Sitemaps were briefly considered as a serialization option
for Resource Maps. Google, Yahoo and Microsoft support the Sitemap
Protocol \cite{sitemaps}, a simple XML file format that allows Web
sites to list the URIs they want crawled by robots. Sitemaps provide for
minimal metadata (e.g., last modification date, update frequency and crawl
priority), but no attempt is made to provide semantic typing, and handling
arbitrary distributed resources is not supported. Indeed, in the interest
of trust, the Sitemap Protocol specifies a significant limitation on URI
paths that can be listed in a Sitemap file. For example, a Sitemap at
level \url{www.foo.com/a/b} can list URIs at level \url{a/b} and below,
but it cannot list URIs at \url{www.foo.com/a/c}, \url{www.foo.com/d/}
or \url{www.bar.com/}.

We made a deliberate decision to avoid the many existing packaging
formats, such as MPEG-21 DIDL \cite{bekaert:mpeg21:ijdl}, METS
\cite{mcdonough:mets}, FOXML \cite{lagoze2006fac}, IMS-CP \cite{ims:cp},
and BagIt \cite{rfc:bagit}.  First, packaging base64-encoded
content in a wrapper document does not resonate well with the
Resource/URI/Representation paradigm of the Web Architecture. Still,
most of these formats also support a by-reference mechanism to deliver
content, in which URIs can be used. However, although these formats
are prominent in their respective communities, they have not gained an
adoption comparable to that of Atom or RDF/XML. And while these approaches
can address identification, and enumeration of distributed resources,
they have uneven capabilities to express the graph-based OAI-ORE model,
due to their hierarchical perspective.

In the course of the OAI-ORE effort, we also attempted to model
aggregations as Atom feeds, not entries \cite{ore:techreport:2008}. We
ultimately decided that was the wrong granularity, especially since common
Web 2.0 reuse scenarios, including use with the Atom Publishing Protocol,
work at the level of Atom entries.  The Atom Syndication Format
was preferred over the various RSS formats in anticipation of using the
Atom Publishing Protocol \cite{rfc:5023}.

Some elements of the POWDER \cite{powder:primer} specifications that
were developed in the same timeframe as OAI-ORE address a problem space
similar to that of OAI-ORE. However, POWDER's focus is significantly
broader, and it approaches the problem from the opposite perspective,

focusing on capabilities to assert (via ``Description Resources'') that
a group of resources share certain properties (e.g. access rights),
rather than asserting arbitrary properties about resources that, for
some reason, are grouped into an aggregation. That is, in POWDER the
notion of shared properties defines an aggregation, whereas in OAI-ORE
an aggregation can be created for any reason deemed important by its
creator. Also, while POWDER provides capabilities to describe a group of
resources using a variety of approaches including regular expressions,
it does not introduce an identity for the aggregation.

\section{Conclusions}

This paper has introduced the OAI-ORE solution to the resource
aggregation problem, which we argue meets a critical need in the
development of cyberinfrastructure and the next generation scholarly
communication infrastructure. By aligning the solution with the
Web Architecture, and by leveraging the practices of the Semantic Web and
Linked Data effort, it will facilitate better integration of scholarly
communication with the mainstream Web, it will make scholarly artifacts
more readily usable with common Web tools and applications, and it will
benefit the broader community by making research materials more visible,
verifiable, and by facilitating unexpected reuse.

While OAI-ORE was motivated by scholarly communication, we believe that
the proposed solution has broader applicability. Aggregations, sets,
and collections are as common on the Web as they are in the everyday
physical world. In many situations it would benefit agents and services
if aggregations were unambiguously enumerated and described, essentially
layering an addition level of resource granularity upon the Web.

Evaluation of the OAI-ORE work depends on its adoption and evolution over
time. The work has so far benefited from significant community involvement
throughout the specification process, and the international team that
developed the solution includes representatives with backgrounds
in scholarly publishing, eScience, repository infrastructure,
digital libraries, Web search engines, linked data, and information
interoperability. Work by early adopters, such as the Foresite project
and John's Hopkins publication workflow project, are promising indicators
that these community contributions have led to a solution that stands
realistic chances for significant adoption.

%ACKNOWLEDGMENTS are optional
\section{Acknowledgments}
This work was supported by the National Science Foundation Divisions
of Information and Intelligent Systems and Undergraduate Education
through grant numbers IIS-0430906, IIS-0643784 and DUE-0840744,
the Andrew W. Mellon Foundation, Microsoft, and the Coalition for
Networked Information. Development of OAI-ORE was based on input from
the OAI-ORE Technical Committee, the OAI-ORE Liaison Group, the OAI-ORE
Advisory Committee, contributors to the OAI-ORE Google discussion
group, and members of the Digital Library Research \& Prototyping
Team of the Los Alamos National Laboratory. Individuals are listed at
\url{http://www.openarchives.org/ore/}.

\bibliographystyle{abbrv}
\bibliography{mln}  % sigproc.bib is the name of the Bibliography in this case
% ACM needs 'a single self-contained file'!
%
% That's all folks!

\balancecolumns % GM July 2000

\end{document}